\documentclass{article}

\usepackage[margin=1in]{geometry}

\usepackage{graphicx} 
\usepackage{subfigure} 

\usepackage{natbib}

\usepackage{algorithm}
\usepackage{algorithmic}

\usepackage{amsfonts, amsmath}

\usepackage{tabu}
\newcommand{\be}{\begin{equation}}
\newcommand{\ee}{\end{equation}}

\begin{document}

\title{Locating recombination hot spots in genomic sequences through the singular value decomposition}

\author{
  Jordan Rodu\\
  Carnegie Mellon University
  \and
  Shane Jensen\\
  University of Pennsylvania
}


\maketitle

\begin{abstract} 
Locating recombination hotspots in genomic data is an important but difficult task.  Current methods frequently rely on estimating complicated models at high computational cost.  In this paper we develop an extremely fast, scalable method for inferring recombination hot spots in a population of genomic sequences that is based on the singular value decomposition.  Our method performs well in several synthetic data scenarios.   We also apply our technique to a real data investigation of the evolution of drug therapy resistance in a population of HIV genomic sequences.  Finally, we compare our method both on real and simulated data to a state of the art algorithm.
\end{abstract} 

\section{Introduction}

There is substantial scientific interest in understanding processes by which viruses evolve in response to medical therapies.  As an example, in the past several decades, researchers have made significant progress in slowing the progression of the human immunodeficiency virus (HIV) which leads to Acquired Immunodeficiency Syndrome (AIDS).  However, the HIV population within a patient can still develop resistance to such therapies, and an understanding of the dynamics of the evolution of that resistance is critical to further progress.

Two mechanisms for genomic evolution via nucleotide sequence changes are mutation and recombination.  A mutation is the identity change of a single nucleotide at a particular location in the genome.  Recombination is the exchange of large portions of a sequence between two copies of the genome contained within each virion.  Recombination events provide an effective opportunity for the virus to achieve a high genetic diversity by combining different blocks of mutations and thereby increasing the possibility that a subset of the population develops resistance to a medical therapy.

Significant effort has gone into understanding the dynamics of HIV evolution by attempting to capture differences in mutation and recombination rates between a treatment HIV population exposed to a therapy, and a control population.  For instance, see \citep{Boni07}, \citep{Husmeier03}, \citep{Schultz09}, \citep{weiller98}, and \citep{Jensen}.

Some previous methods attempt to estimate the phylogenetic trees associated with a population of genomic sequences.  These are model-based methods that would ideally reconstruct, along with their probabilities, all possible ancestral histories of the sequence population.  The possibility of recombination presents a combinatorial challenge for these methods, as recombinations necessitate the construction of ancestral recombination graphs (ARGs).  Markov models are often employed in conjunction with phylogenetic trees infer recombination rates without estimating ARGs \citep{LiSte03}. 

There are alternative approaches that not model-based, but instead focus on establishing a similarity between each sequence in the population based on an accepted distance metric.   A potential recombination location can then be inferred by looking for locations in which particular sequences are close on one side of the proposed recombination location, and far on the other side.  For instance, \citep{weiller98} uses a sliding window to compare the Hamming distance between each sequence and every other sequence in the population and measures the correlation between the comparisons before and after a proposed recombination location.  High correlation indicates no recombination, while low correlation indicates the possible presence of a recombination, since the target sequence relates to the other sequences differently before and after the location of interest.

In general, these distance-based methods can be orders of magnitude faster than their model-based counterparts, but they do come with some drawbacks.   First, graphical examination of the results is often needed to determine recombination location.  Secondly, they often seek to detect recombination on a sequence-by-sequence basis, comparing each sequence against the rest of the population sequentially, rather than inferring recombination simultaneously across all sequences.  

In this paper, we present a non-model-based approach that addresses the issues described above for distance-based methods.   Specifically, we provide an automated procedure for locating recombination hot spots based on the simultaneous analysis of an entire sequence through use of the Singular Value Decomposition (SVD).   Although we also use the Hamming distance, our approach contrasts \citep{weiller98} by detecting recombination locations that are \emph{hot spots} characteristic of the entire population, rather than recombination locations based on comparing a particular sequence to the rest of the population.  This enables us to borrow strength from the entire population of sequences when inferring recombination locations, thereby avoiding  some of the noisy interpretations of a sequence by sequence analysis.

We compare our results with a state of the art algorithm, RECCO \citep{RECCO}.  RECCO analyzes potential recombination locations by analyzing the cost of mutations between two sequences by introducing a recombination breakpoint.  We show that our SVD based method exhibits superior performance in locating recombination hot spots and is a valuable addition to the genomic toolbox.

Our paper is organized as follows.  Section \ref{sec:method} outlines the mathematical foundations of our SVD method, as well as a full description of our approach.  In Section \ref{sec:simulation}, we explore the precision and recall of our method under several synthetic data scenarios.  We then apply our SVD recombination detection method to a real dataset consisting of genomic sequences from HIV populations before and after the application of a drug therapy in Section \ref{sec:real}. 

\section{SVD-based Recombination Detection}
\label{sec:method}

Genomic sequence data is typically organized as a set of long character strings where the character alphabet is limited to a small number of characters (four unique characters in the case of RNA or DNA sequences).   When describing the evolution of these genomic sequences, we are interested in two mechanisms for sequence differences: the location of mutations and the locations of recombination hot spots.  

\subsection{Hamming Distance}

For both of these characteristics, we need to work not with the sequences themselves but rather where the sequences differ. 
This is clear in the case of mutations.  To see this is true for recombinations, let $t$ be the location of a recombination hot spot in a population of sequences.  Consider two sequences, one that is entirely a replica (with some random mutations) of parent $A$, and one that is a replica of parent $A$ up until $t$, and a replica of a different parent $B$ afterwards.  When we compare the two sequences, they effectively agree with each other up to position $t$, and, provided parent $A$ and $B$ differ in a number of mutations after the recombination, our two sequences disagree after $t$.

To make this mathematically rigorous, let the hamming distance for two subsequences be the number of locations where the two subsequences differ.  Letting $\mathbb{I}_x$ be $1$ if event $x$ occurs, and zero otherwise, we have that the hamming distance over two subsequences $a$ and $b$ is $\mathrm{HAMM}(a, b) = \sum_k \mathbb{I}_{a_k != b_k}$ where $k$ ranges over the number of positions in strings $a$ and $b$ (the strings are assumed to be of equal length).

For a pair of sequences $s_i$ and $s_j$, we can obtain a smoothed hamming distance sequence by defining a window of length $w$, and at each position $t$, calculating the hamming distance between the positions $t-w$ and $t+w$, associating this distance with time $t$ in the smoothed Hamming distance sequence.

Formally, let $s_j$ be the $j^\mathrm{th}$ sequence in a population of sequences, $s_j[y]$ be the $y^\mathrm{th}$ entry in $s_j$, and $s_j[y:z]$ be the subsequence of sequence $s_j$ corresponding to entries $y$ through $z$.  Define a new sequence $x_{i, j}$, the smoothed Hamming distance sequence, associated with sequences $s_i$ and $s_j$ such that, for window size $w$ we have
\begin{align*}
x_{i,j}[y] = \mathrm{HAMM}(s_i[y-w:y+w], s_j[y-w:y+w])
\end{align*}

Finally, let $X$ be the matrix whose rows are the sequences $x_{i, j}$.  The rows of this matrix $X$ are the smoothed Hamming distances for all sequence pairs in a population.

\subsection{The Singular Value Decomposition}
The singular value decomposition (SVD) is a matrix decomposition that decomposes a matrix $A$ into three components, $U$, $D$, and $V$ where $A = UDV^\top$ where $U$ and $V$ are orthonormal matrices, so $U^\top U=I$ and $V\top V=I$, and $D$ is a diagonal matrix with diagonal elements $D_{ii}\geq 0$.  Columns of $U$ and $V$, $u_i$ and $v_i$, are called the left singular vectors and right singular vectors, respectively, and the diagonal elements of $D$ are called singular values.  We assume that the singular values are ordered, so $D_{11} > D_{22} > \ldots > D_{kk}$.  Alternatively, we can express the SVD as the sum of rank $1$ matrices, so $A = \sum u_i D_{ii} v_i^\top$.  Because of the ordering we place on the singular values, then $u_1 D_{11} v_1^\top$ is the best rank $1$ approximation to $A$, and $A=\sum_{i=1}^k u_i D_{ii} v_i^\top$ is the best rank $k$ approximation to $A$.

With this interpretation in mind, we can think of the first right singular vector as providing a line of best fit to the rows of our matrix, with the first left singular vector scaling the right singular vector and possibly flipping the signs of the entries of the first right singular vector.   The second pair of singular vectors can be seen as the same operation on the residuals from the first singular vector pair.  The remaining singular vectors can continue to be interpreted in the same way.

\subsection{Applying the SVD to recombination hot spots}

We use this interpretation of the SVD in our analysis of recombination hot spots.  We begin with our matrix $X$ as defined above.  Recall that a row in this matrix represents the smoothed hamming distance sequence between two sequences in the population.  The value at an entry in this row is the hamming distance between subsequences of the sequences centered around that entry.

The first singular vector pair captures the major mutation profiles and won't be discussed here besides the following two points.  First, we will think of the first singular vector pair as essentially centering (after scaling) the smoothed hamming sequences, and second, the entries in the first singular vector pair are all positive because entries in $X$ are all greater than or equal to $0$. This is not true for the rest of the singular vector pairs.  

\begin{figure}[tbph]
\vspace{0in}
\begin{center}
\includegraphics[width=.49\linewidth]{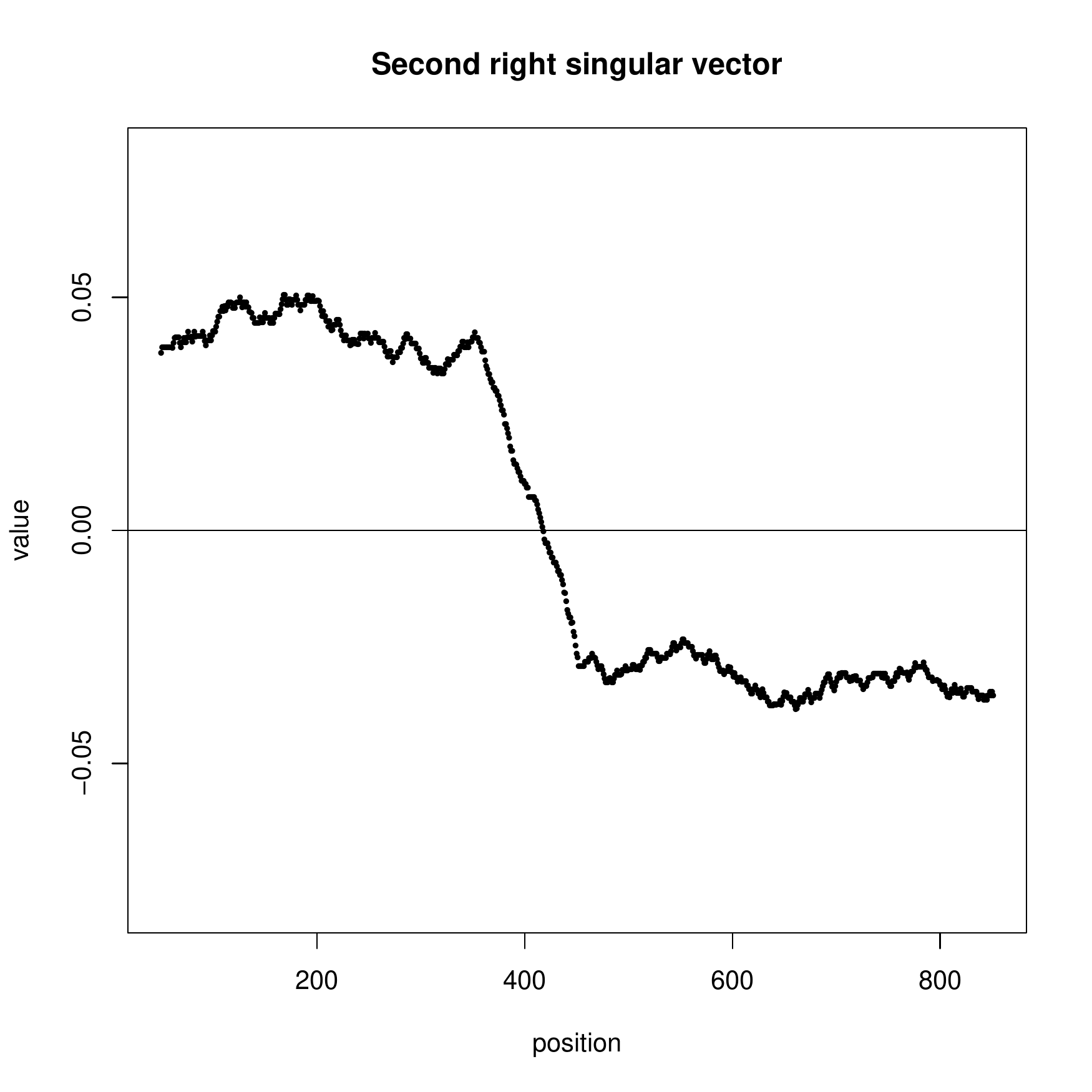}
\includegraphics[width=.49\linewidth]{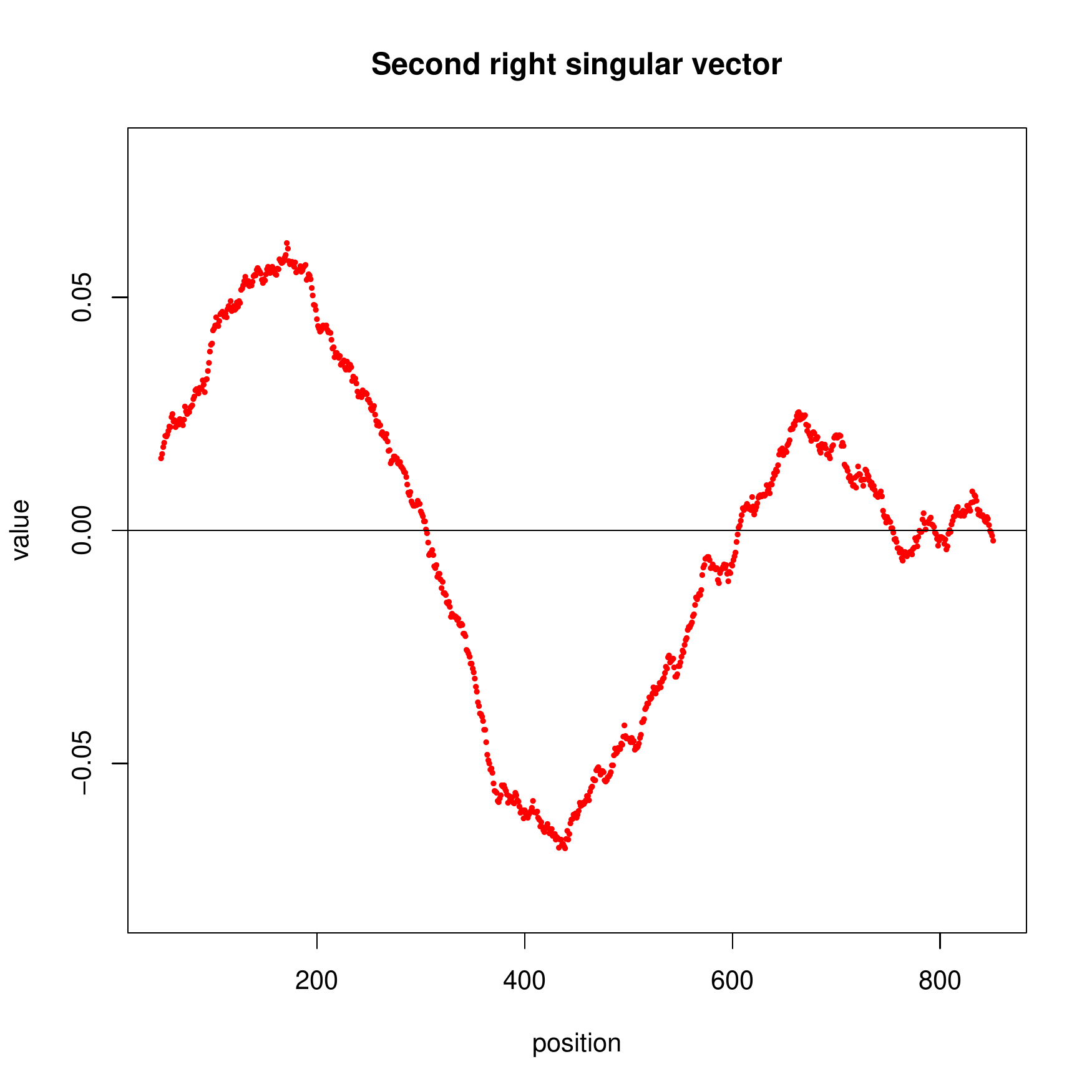}
\caption{Entries of second right singular vector, with recombination (left), and without recombination (right).}
\label{rsv_unsorted}
\end{center}
\end{figure}

We will use the second singular vector pair, and in particular the second right singular vector, to locate recombination hot spots.  As previously mentioned, we can think of the first singular vector pair as centering the sequences.  The key is that in smoothed hamming sequences in which the compared sequences come from similar parents on one side of a recombination location and not from similar parents on the other side of the recombination location, the first singular vector pair will have underestimated the entries on the side where the parents are not similar parents, and will have overestimated the entries on the side where the parents are similar parents.  This trend in the residuals is the trend captured in the second singular vector pair (see \ref{rsv_unsorted}).  Under this scenario, we expect the right singular vector to be positive (or negative) before the recombination location, and switch to negative (or positive) afterwards.  Because we are working with smoothed hamming distance sequences, this transition will be gradual, and the transition will be centered roughly around the recombination location.

Some rows will not load onto the second right singular vector at all (specifically the rows whose compared sequences have similar parents the entire length of the sequence), some will load positively onto the second right singular vector, and others will load negatively (this stems from the fact that in the population there are compared sequences with shared parents before the recombination and not after, as well as compared sequences with shared parents after the recombination and not before).

We want to first determine if there exists a recombination hot spot in a population, and second infer its location.  We have already seen clues as to how to determine the location of a recombination hot spot.  We will be looking for a location in the second right singular vector that signifies a major change from mostly positive entries to mostly negative locations (or, equivalently, from mostly negative to mostly positive).  To characterize this significant change we look to the slope of successive entries in a window pre-determine size. However, we must first determine whether or not a recombination hot spot is even present in a population, which we do by examining the average distance between successive entries in the second right singular vector.   

\subsection{Inferring the existence of recombination}

Letting $N$ be the length of $x$, we define the \emph{mean difference} between successive entries in $x$ is
\begin{align*}
\overline{d}(x) &= \sum_i \frac{|x[i] - x[i+1]|}{N}
\end{align*}
We say a population has a recombination if $\overline{d}(x) < \gamma_x$ for some threshold $\gamma_x$ that depends on the population.

$\gamma_x$ is set by appealing to simulated populations without a recombination, created by permuting the entries of the gene sequences (where for each simulated population there is a single permutation applied to each sequence).  Recall that in the case of a recombination some pairs of sequences have a higher rate of disagreement on one side of the recombination as compared to the other side of the recombination.  Permuting the entries has the effect of redistributing the locations in which pairs of sequences disagree, simulating a population with no recombination event.

\begin{figure}[tbph]
\vspace{0in}
\begin{center}
\includegraphics[width=.49\linewidth]{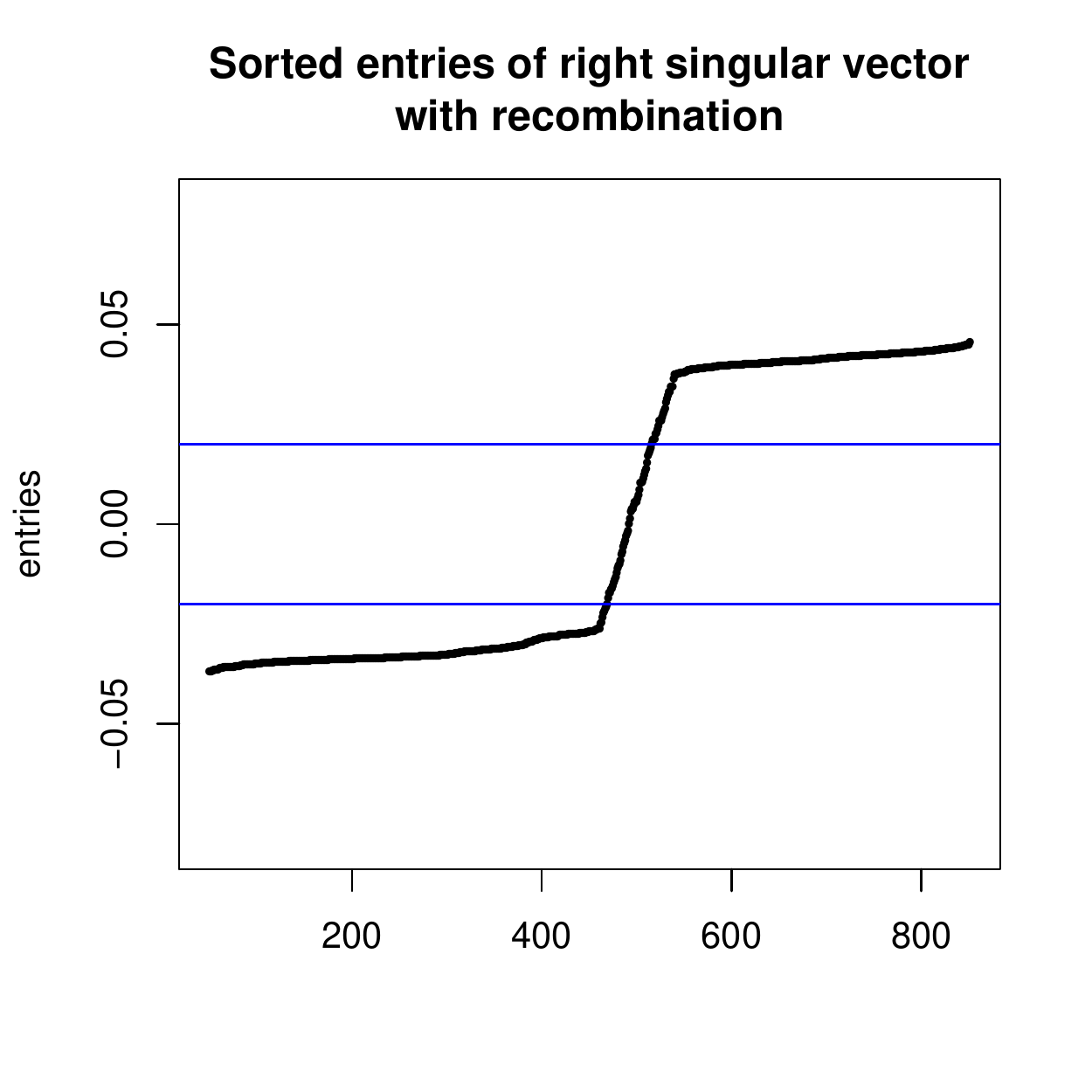}
\includegraphics[width=.49\linewidth]{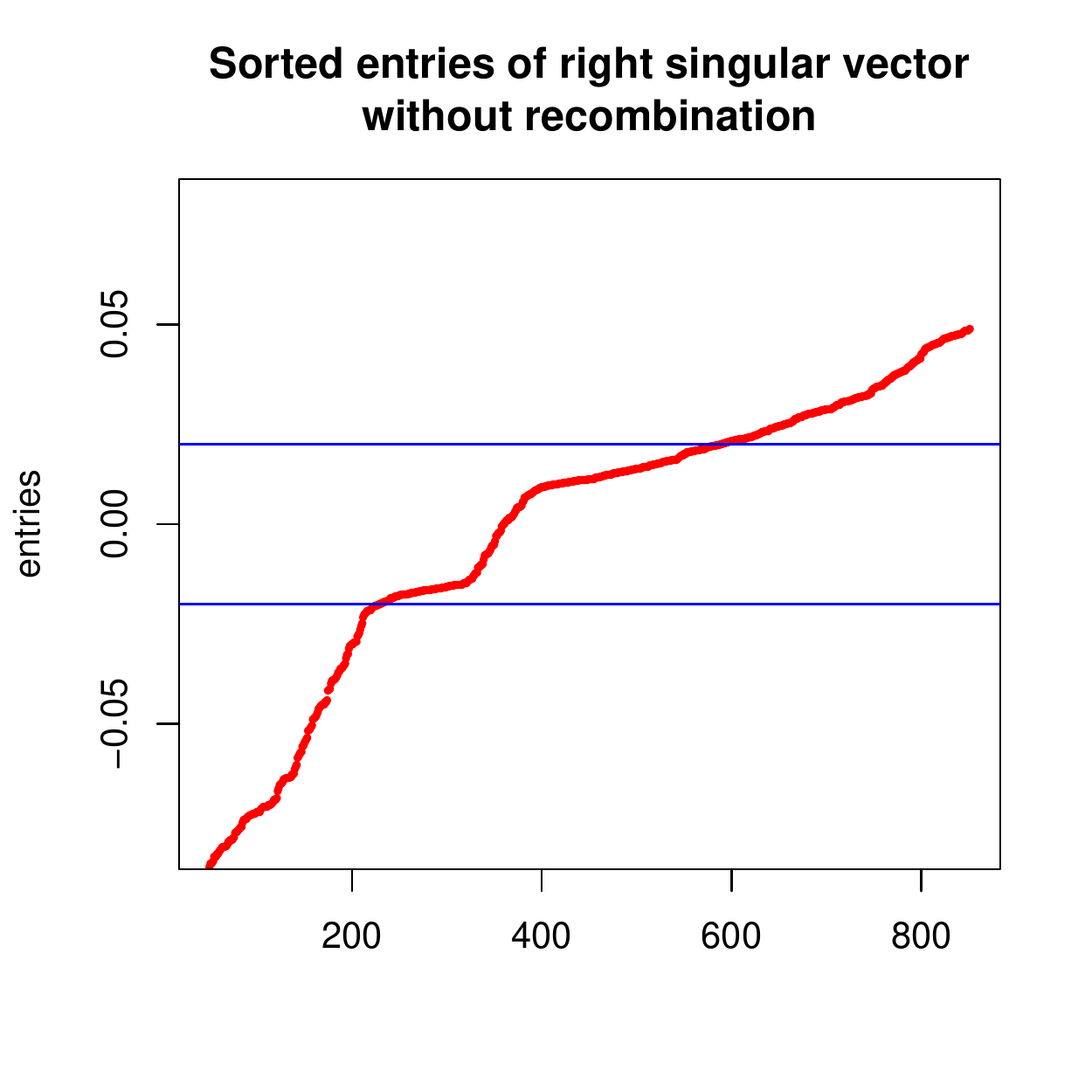}
\caption{Sorted entries of second right singular vector, with recombination (left), and without recombination (right), the blue bands are at $-.02$ and $.02$.}
\label{rsv_sorted}
\end{center}
\end{figure}

There are a few things to unpack to understand why the mean method is highly successful for identifying the presence of a recombination.  Recall that the Euclidean norm of the singular vectors is 1 (and in fact letting $V$ be the matrix of right singular vectors as above, we have $V^\top V = I$ where $I$ is the identity (with appropriate dimension).  The entries of a second right singular vector in the case of a population with a recombination event tend to start positive or negative, and switch only when there is a recombination event.  The entries do not usually vary within a small window, and in general, the entries spend very little time near zero, usually coming close to zero only near the recombination event location.  This can be seen in the left panel of figure \ref{rsv_sorted}.

On the other hand, the right singular vector corresponding to a population with no recombination wanders quite a bit more within a small window, with a larger proportion of entries closer to zero.  Because more entries are located close to zero than in the recombination case, the entries that are large in magnitude have to be comparatively larger than in the recombination case, to compensate.  Further they cover that distance more often.  In the recombination case, entries to the left of the recombination location tend to cluster around a positive (or negative) magnitude, and then switching to concentrative around a negative (resp. positive) magnitude after the recombination location.  Intuitively, not only are they not covering a wide range of magnitudes in general, but they aren't straying very far one entry to the next, except when they travel because of the recombination event.  In the case of no recombination populations, the entries are freely wandering from positive to negative entries, so we expect the absolute differences between successive entries to be larger.

\subsection{Inferring the locations of recombination}

To identify the recombination location, we want to identify a subsequence of the second right singular vector that captures the slope of the transition effect.  Because, as discussed above, the transition effect occurs over a window of size $w$, we obtain the ``running slope'' in blocks of size $w$ of the second right singular vector.  The window that contains the maximum slope is inferred to be the window containing the transition effect.

At present we consider two different notions of slope.  One is to take the absolute difference between the first and last points in a window.  As the second singular vector tends to be smooth in the presence of a single recombination event, this is a decent proxy for slope.  The second is to simply take the ordinary least squares of the entries in each window.  

More formally, for each entry $i$ in the second right singular vector $x$ associated with some population, we assign a slope according to the \emph{diff} method as follows:
\begin{align*}
\mathrm{diff}_w(x[i]) &= \left|x[i-\frac{w}{2}] - x[i + \frac{w}{2}]\right|
\end{align*}
where $w$ is the window size used to determine the ``moving'' hamming distance for the sequence pairs.

In the case of the ordinary least squares method (OLS), letting $y_i = x[i-\frac{w}{2}, i+\frac{w}{2}]$, and $z$ be the vector from $-\frac{w}{2}, \frac{w}{2}$ in unit increments, we assign a slope to each entry in $x$ as given by
\begin{align*}
\mathrm{ols}_w(x[i]) &= |<y_i, z>|
\end{align*}
where $<\cdot, \cdot >$ is the dot product between two vectors.

We propose inferring a recombination hot spot at the location corresponding to the slope of greatest magnitude.

\subsection{Inferring multiple recombination locations}
We have focused on inferring the presence and location of a single recombination hot spot.  Our algorithm, however, applies to an arbitrary number of hot spots.  To infer the presence of a second recombination hot spot, instead of looking at the second right singular vector, we look at the third right singular vector.  Analysis proceeds exactly as before.


\section{Evaluation with Synthetic Data}\label{sec:simulation}

We consider the following synthetic data scenarios.   We simulated datasets with either $0$, $1$, or $2$ recombination hot spots.   For each dataset, we generate a population of $100$ sequences (each of length 1000) using the following scheme. 
First, two parental sequences are generated by starting with two identical sequences and randomly mutating each position of those two sequences independently at a rate $r_c$, which we call the \emph{common mutation rate}.    If there is no recombination present in a trial, $100$ sequences are sampled from these parent sequences.  

If there does exist a recombination hot spot, two recombinant daughter sequences are generated, where daughter $1$ is identical to parent $1$ up to the recombination location and identical to parent $2$ after the recombination location.  Daughter $2$ is identical to parent $2$ up to the recombination location, and parent $1$ after the recombination location. $100$ sequences are then randomly sampled from this collection of two parents and two daughters.  Datasets with two recombination locations are handled similarly.  We can think of a sequence in a two recombination case as being a concatenation of three smaller subsequences, and to generate the daughters of the population we consider all possible combinations (i.e. AAA, AAB, ABA, ABB, BAA, BAB, BBA, BBB, where A and B indicate the parent origin for that particular subsequence).

Finally, for all datasets (with and without recombinations), mutations are then added to random positions of each sequence within the 100 sequence population at a rate $r_i$, which we call the \emph{individual mutation rate}.    We consider common mutation rate values of either $.05$ (low) or $.25$ (high), and individual mutation rate values of either $.05$ (low) or $.25$ (high).   We generated synthetic datasets under all combinations of these parameters, with 20 datasets generated for each combination. 

For each synthetic dataset, we used our SVD method to estimate the number of recombination hot spots as well as the locations of those recombination hot spots (if they were estimated to exist for that dataset). 

Table \ref{tab:nori2} gives the false positive (FP) and false negative rates for our detection of the true number of recombination hot spots under each of our synthetic data scenarios.

\begin{table}[ht]
\caption{False positive (FP) and false negative (FN) rates for identifying presence of a recombination.  Com/Ind represents the combination of the low vs. high value of the common mutation rate vs. individual mutation rate}
\begin{center}
\begin{tabular}{c|c|c|c|c|c|c|}
  \cline{2-7}
  & \multicolumn{6}{|c|}{Number of True Recombinations}\\ \cline{2-7}
  & \multicolumn{2}{|c|}{0} & \multicolumn{2}{|c|}{1} & \multicolumn{2}{|c|}{2}\\ \cline{1-7}
  \multicolumn{1}{|c|}{{\bf Com/Ind}} & {\bf FP} & {\bf FN} & {\bf FP} & {\bf FN} & {\bf FP} & {\bf FN} \\ \cline{1-7}
  \multicolumn{1}{|c|}{low/low} & .01 & NA & .03 & 0 & .045 & .25 \\ \cline{1-7}
  \multicolumn{1}{|c|}{low/high} & 0 & NA & .04 & .30 & .065 & .5 
  \\ \cline{1-7}
  \multicolumn{1}{|c|}{high/low} & .01 & NA & 0 & 0 & .045 & 0 \\ \cline{1-7}
  \multicolumn{1}{|c|}{high/high} & .02 & NA & .06 & 0 & .025 & .195 \\ \cline{1-7}
  \end{tabular}
\end{center}
\label{tab:nori2}
\end{table}

Overall, our SVD method performs quite well in terms of identifying the true number of recombinations.   Performance suffered in the case of a low common mutation rate and high individual mutation rate in detecting the correct number of recombinations.  This data scenario is particularly difficult as we have a high rate of individual mutations, which introduces additional noise into our data.    A low rate of common mutations is disadvantageous since common mutations are going to be harnessed by our method to narrow down the location of potential recombination events.  When the common mutation rate is low, then there are large gaps between mutations.    If there is is a true recombination hotspot at location 500 but if no sequence in the population exhibits a mutation between, say, entries 400 and 600, a recombination anywhere between 400 and 600 would yield the exact same population, making it impossible to determine that the true recombination location is near 500.  If the common mutation rate were higher, and say the no-mutation buffer around (the true recombination) entry 500 were only about 5 or 10 entries, then we can expect a good algorithm to often get decently close to the truth.

\begin{figure}[tb]
\vspace{0in}
\begin{center}
\includegraphics[width=.45\linewidth]{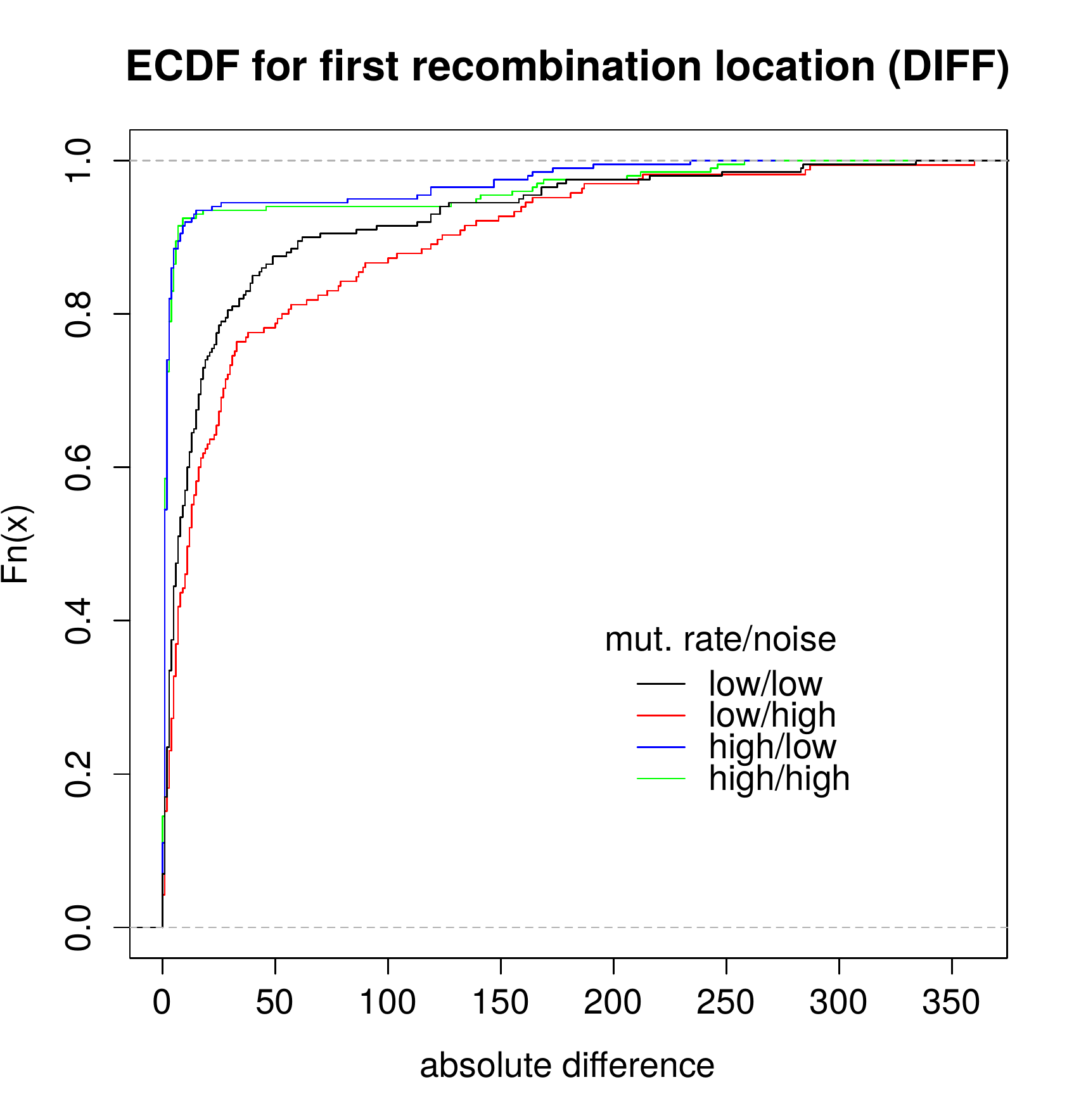}
\includegraphics[width=.45\linewidth]{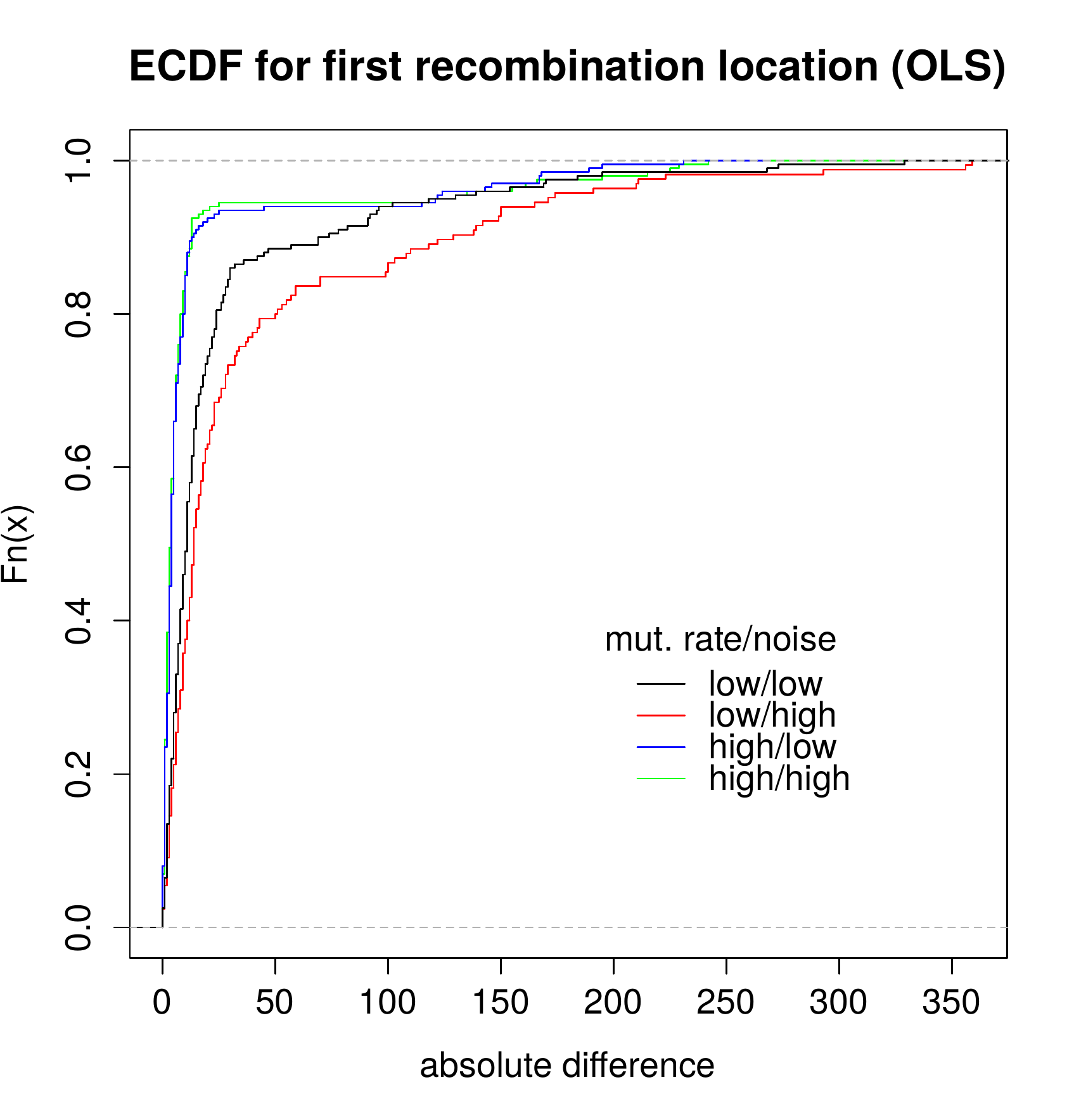}
\caption{Empirical CDF for absolute difference between inferred recombination location from the second right singular vector and closest true recombination location for the {\bf left:} Diff method, and {\bf right:} OLS method.}
\label{ecdfs}
\end{center}
\end{figure}

\begin{figure}[tb]
\begin{center}
\includegraphics[width=.4\linewidth]{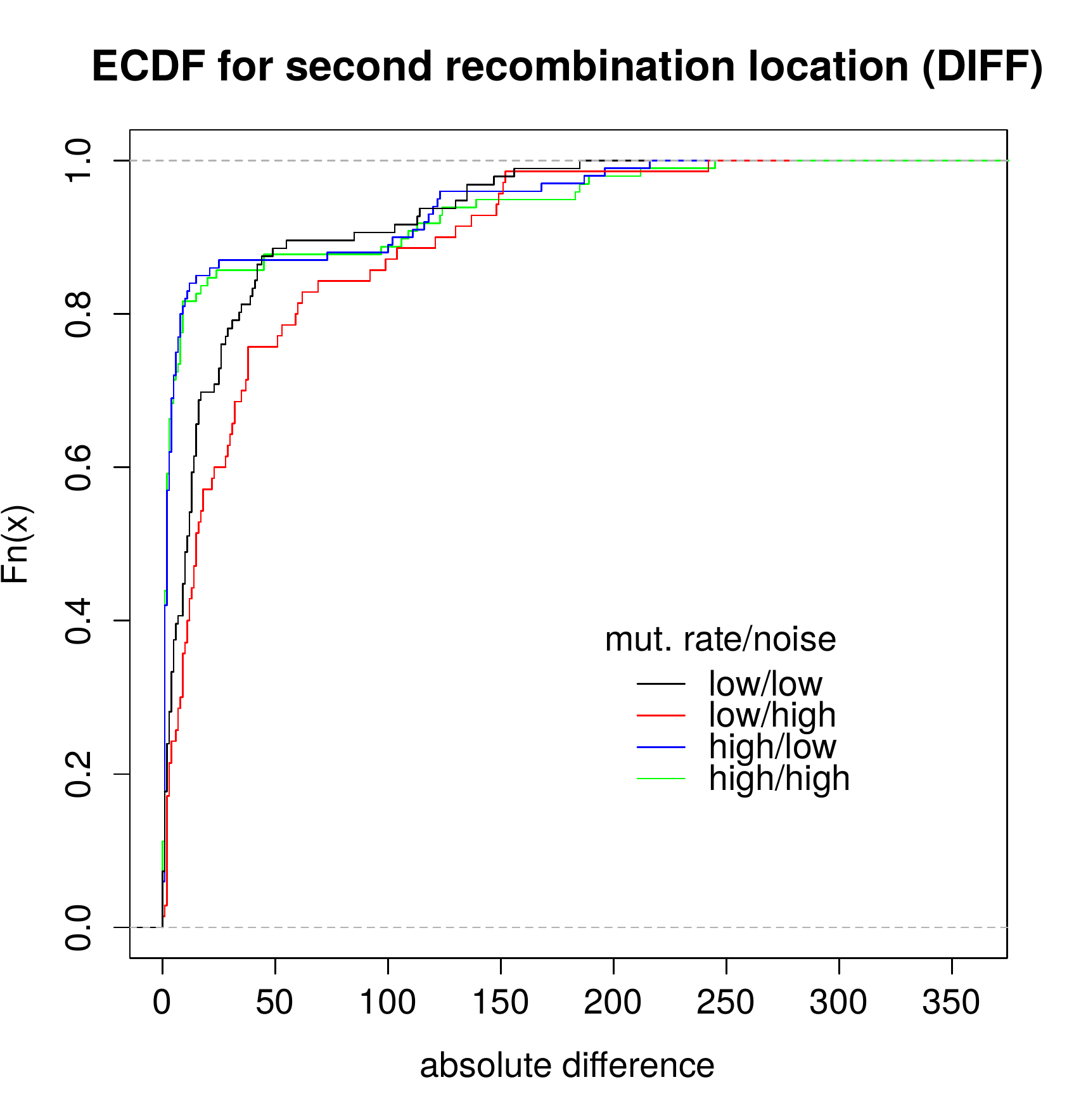}
\includegraphics[width=.4\linewidth]{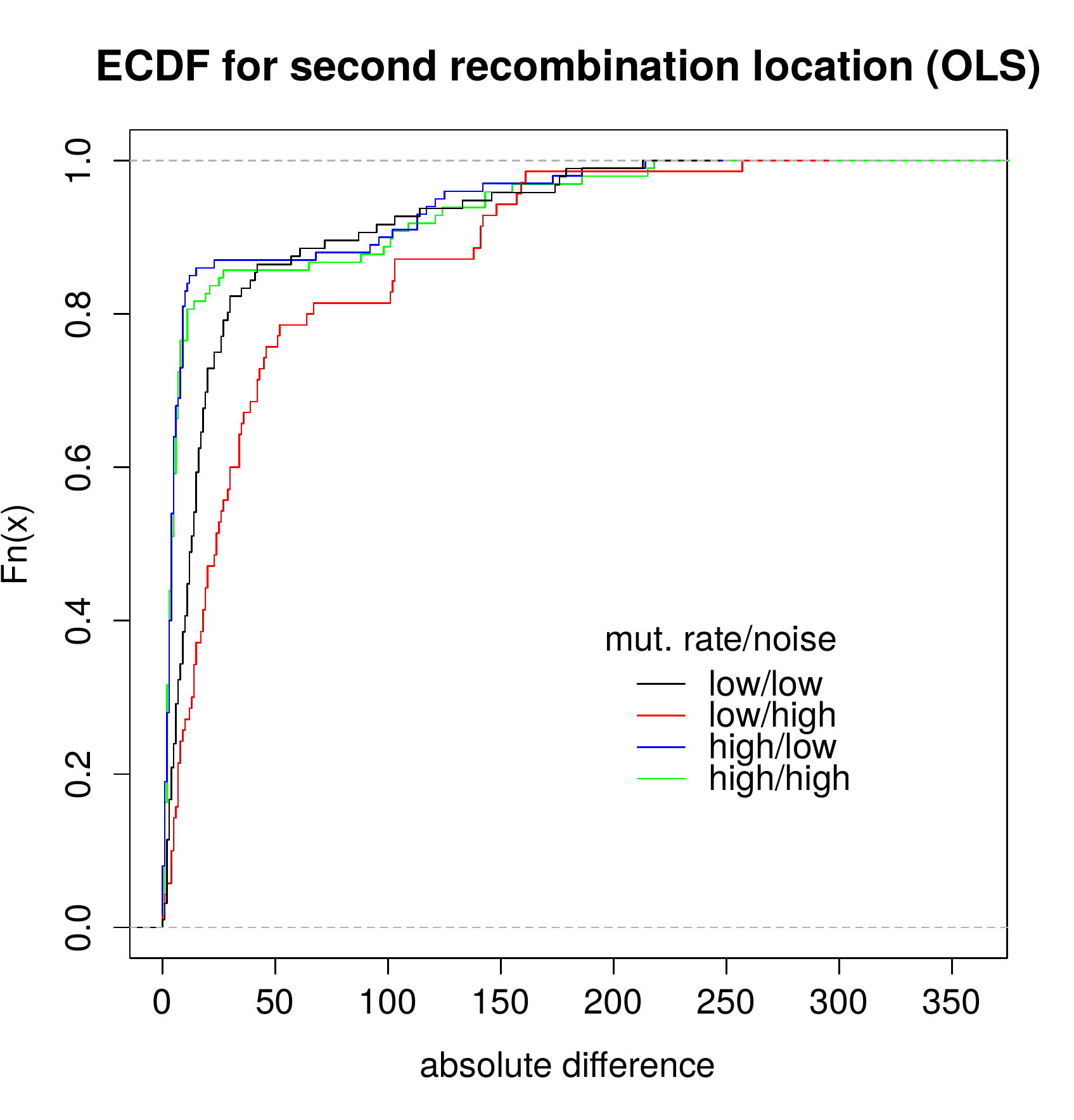}
\caption{Empirical CDF for absolute difference between inferred recombination location from the second right singular vector and closest true recombination location for the {\bf left:} Diff method, and {\bf right:} OLS method. }
\label{ecdfs2}
\end{center}
\end{figure}

For those scenarios in which the correct number of recombinations were inferred, we can also evaluate our SVD method in terms of the accuracy of the estimated recombination locations.   In Figures \ref{ecdfs} and \ref{ecdfs2}, we present the empirical CDF of the distance between the true recombination location and the second right singular vector (Figure \ref{ecdfs}) and the third right singular vector (Figure \ref{ecdfs2}).   For simulations with only a single recombination location this is simply the distance between the inferred location and the true location.  For simulations with two recombination locations, this is the distance between the inferred recombination location and the closest true recombination location.  We present the empirical cdfs for both the Diff method and the OLS method.

Since our procedure infers the best location by seeking out the slope with maximal magnitude, it is possible for the optimal location in the third right singular vector to be inferring the same location as the optimal entry in the second right singular vector.  Indeed this happens roughly $10\%$ of the time in our simulations.  For most cases, however, this is not an issue, and the inferred location in the third right singular vector is \emph{not} the same location found in the second right singular vector.  

The OLS and Diff methods both provide excellent inference on recombination locations, both inferring a recombination location within a small distance of the true recombination location with high probability.  Of the inferred locations falling within a small distance of the true recombination location, OLS locations tend to be in a tight region around the truth, while DIFF locations tend to be slightly more spread out within that small region. On the other hand when inferred locations fall outside the small region, OLS inferred locations seem to fail more dramatically than the DIFF locations.



\subsection{Comparison to RECCO}
Finally, we compare our performance on synthetic datasets with the the performance of current state of the art recombination detection procedure RECCO \citep{RECCO}.  We consider sequences with zero true recombination locations and one true recombination location to evaluate, as above, the number of recombination hotspots inferred, and to assess the accuracy of the inferred recombination locations. For simplicity, we use only the DIFF method for comparison to RECCO.

RECCO attempts to infer recombination locations by proposing, per sequence, a recombination in locations that substantially reduce the number of mutations.  In order to guard against arbitrarily many recombinations, a penalty is imposed for introducing a recombination.  The net savings at each location is compared to net savings from position-permuted datasets that induce a null distribution over net savings.  



In order to infer hotspots using RECCO, we use the breakpoint p-values for the data set and collect locations with dataset p-values $\leq .05$.  As with all methods, depending on the mutation rate it is difficult to narrow possible recombination locations down to a single position.  The measure most closely related to the number of inferred hotspots, therefore, is the number of contiguous regions of sequence locations with small p-vales.  In defining a contiguous region, it is possible to have a sequence of locations with p-values $\leq .05$, with the exception of a few isolated locations.  In order to avoid splitting such regions into many more smaller regions, we only split a regions if a gap exists of more than $4$ locations.  Results on false positive and negative rates for RECCO are presented in table \ref{RECCOfp}, along with the corresponding rates for our SVD method, for convenience.


\begin{table}[ht]
\caption{False positive (FP) and false negative (FN) rates for identifying recombinations through the number of contiguous sequences of small p-values.  Relevant rates from the SVD method are provided from \ref{tab:nori2} for convenience.}
\begin{center}
\begin{tabu} to .4\textwidth {X[c]|X[c]|X[c]|X[c]|X[c]|X[c]|X[c]|}
  \cline{2-7}
  & \multicolumn{6}{|c|}{Number of True Recombinations}\\ \cline{2-7}
  & \multicolumn{2}{|c|}{0} & \multicolumn{4}{|c|}{1}\\ \cline{2-7}
  & \multicolumn{1}{|c|}{RECCO} & \multicolumn{1}{|c|}{SVD} & \multicolumn{2}{|c|}{RECCO} & \multicolumn{2}{|c|}{SVD}\\ \cline{1-7}
  \multicolumn{1}{|c|}{{\bf Com/Ind}} & {\bf FP} & {\bf FP} & {\bf FP} & {\bf FN} & {\bf FP} & {\bf FN} \\ \cline{1-7}
  \multicolumn{1}{|c|}{low/low} & 3.09 & .01 & 4.61 & .02 & .03 & 0 \\ \cline{1-7}
  \multicolumn{1}{|c|}{low/high} & 9.77 & 0 & 17.35 & 0 & .04 & .30 
  \\ \cline{1-7}
  \multicolumn{1}{|c|}{high/low} & 3.09 & .01 & 3.68 & 0 & 0 & 0 \\ \cline{1-7}
  \multicolumn{1}{|c|}{high/high} & 10.19 & .02 & 17.64 & 0 & .06 & 0 \\ \cline{1-7}
  \end{tabu}
\end{center}
\label{RECCOfp}
\end{table}




As can be seen from the table, RECCO has a tendency to infer a substantial number of contiguous regions of low p-value sequence locations, both in the null case and in the one recombination case.  The biggest concern with this is when these contiguous regions are spread across a large portion of the sequence, rendering RECCO somewhat ineffective in pinpointing a fairly targeted region of interest.

\begin{figure}[tbph]
\vspace{0in}
\begin{center}
\includegraphics[width=.5\linewidth]{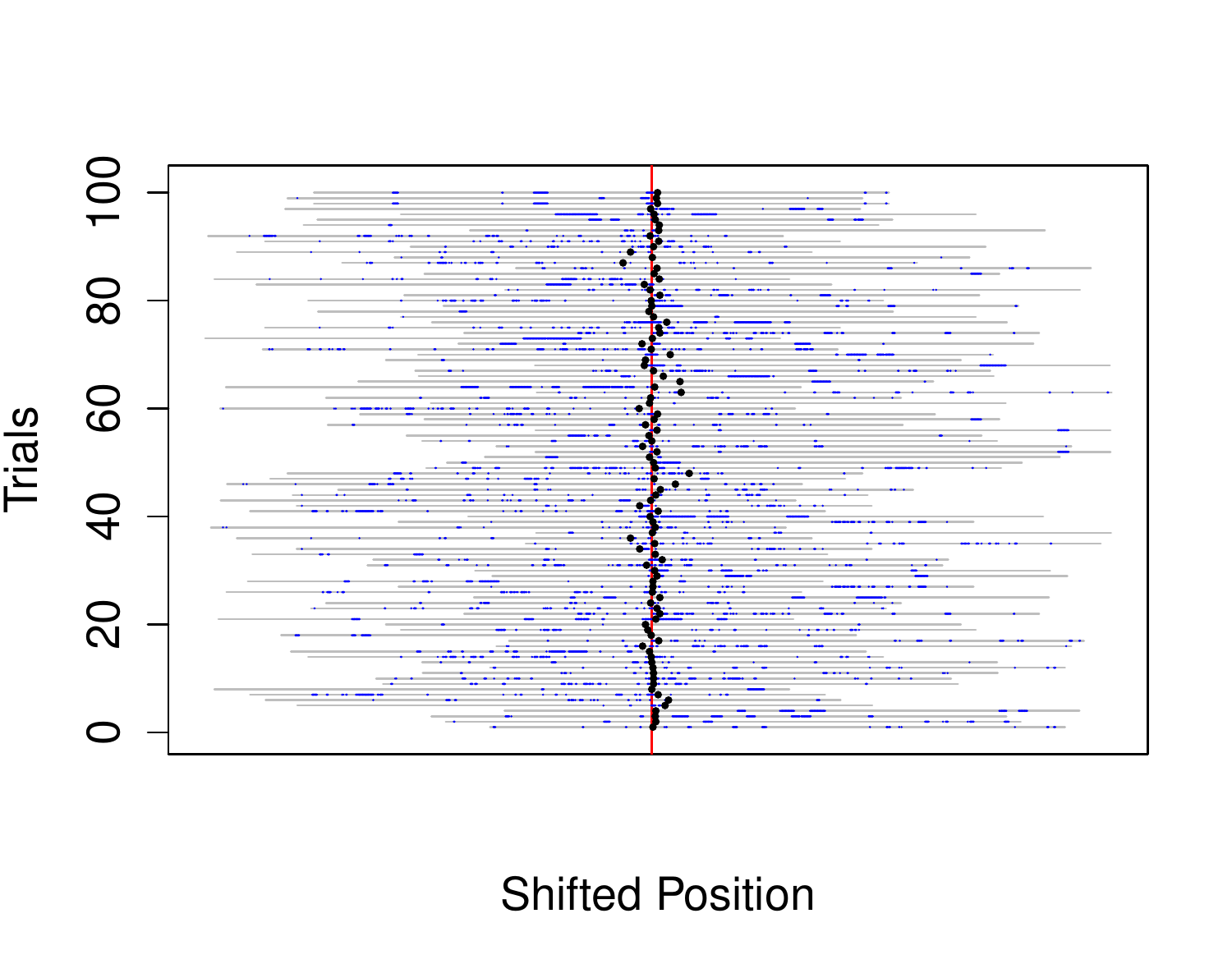}
\caption{Inferred recombination hot spots on simulated trials.  Positions shifted to align true recombination locations (red line).  SVD method inferred position in black, RECCO in blue}
\label{shiftedRECCO}
\end{center}
\end{figure}

The ineffectiveness of RECCO in specifying a narrow region of interest is highlighted in figure \ref{shiftedRECCO}.  Among trials with one true recombination hotspot, we randomly select 100 trials among all $4$ ``Com/Ind'' mutation profiles.  Each trial is represented in the plot by a grey horizontal line.  For ease of presentation, the grey lines have been horizontally shifted relative to one another in order to align the recombination hotspots, represented in the plot with a red vertical line.  The recombination location inferred by our method is indicated with a black dot, and the blue dots indicate regions of small p-values inferred by RECCO.  

\section{Application to HIV Genomic Sequences}\label{sec:real}

We use our SVD method to detect recombination hot spots in HIV populations from patients that were undergoing a drug therapy.    Our sequence data consists of the genomic sequence that codes for the envelope ({\it env}) protein of HIV.   In our dataset, we have 21 {\it env} sequences prior to drug administration, and 24 {\it env} sequences after drug administration.  HIV can develop resistance to the drug through even a small number of mutations, and recombination provides a vehicle for the sharing of advantageous mutations between HIV virions.  

We applied our SVD method in three different analyses:  1. treating the pre-drug sequences as a single population, 2. treating the post-drug sequences as a single population and 3. considering the combined the pre-drug and post-drug sequences as a single population.

\begin{figure*}[tb]
\vspace{0in}
\begin{center}
\includegraphics[width=.48\linewidth]{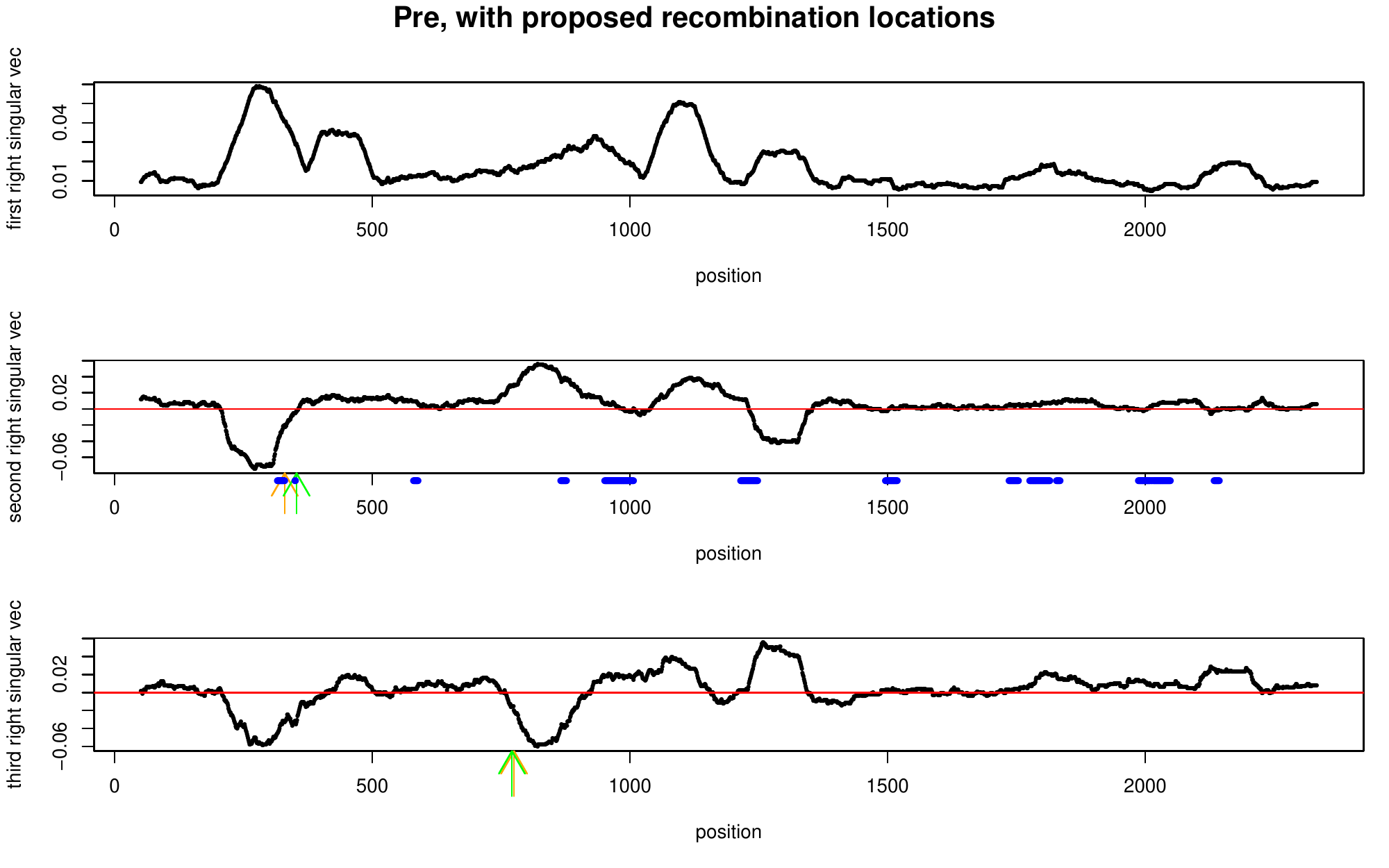}
\includegraphics[width=.48\linewidth]{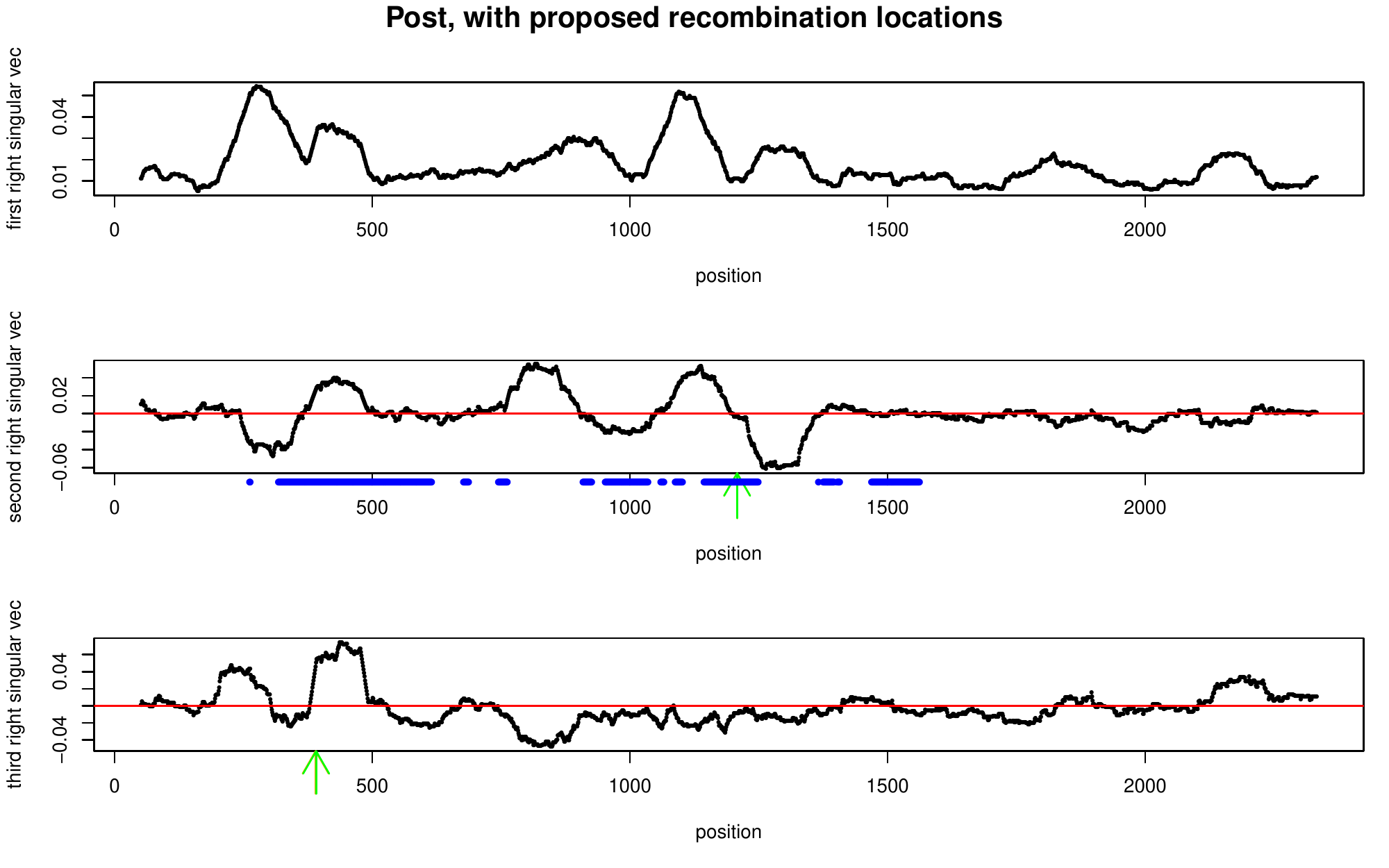}
\includegraphics[width=.48\linewidth]{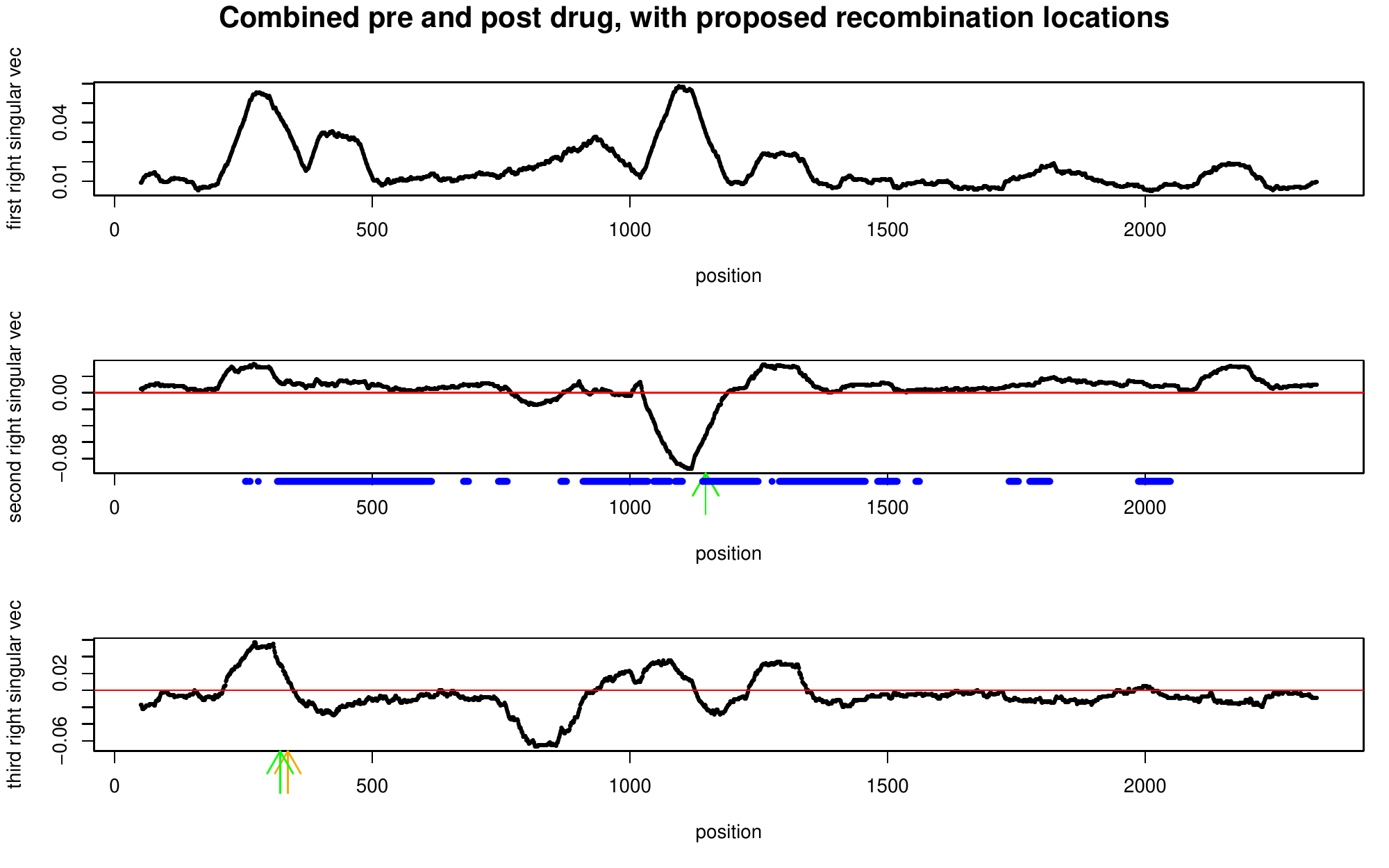}
\caption{First three right singular vectors from our method applied to pre-drug (top left), post-drug (top right) and combined pre-drug and post-drug (bottom left) population of sequences.  The orange arrows represent the recombination location suggested by the least squares method, and the green arrows represent the recombination locations suggested by the diff method (with them agreeing in cases were there is no orange arrow).  The red lines indicate $y=0$.  For each set of sequences we specify the RECCO regions of interest by blue dots just underneath the plots of the second right singular vectors}.
\label{rightprepostcombined}
\end{center}
\end{figure*}

Figure \ref{rightprepostcombined} (top left) shows the first three right singular vectors from our method applied to pre-drug population of sequences.   The blue arrows represent the recombination location suggested by the least squares method, and the green arrows represent the recombination locations suggested by the diff method.  Figure \ref{rightprepostcombined} (top right) shows the first three right singular vectors from our method applied to post-drug population of sequences.  Figure \ref{rightprepostcombined} (bottom)  shows the first three right singular vectors from our method applied to the combined pre-drug and post-drug population of sequences.  

Although the first right singular vector does not provide any inference in terms of recombination, it does allow for insight into the mutation structure of the genomes.   The first right singular vector (the mutation profile) remains consistent across all three sequence populations.  The highest peak in the first right singular vectors is around positions 300-400 of the {\it env} sequence, with roughly corresponds to the location of the V1/V2 loops, which is known to be a highly variable region of the HIV genome  \citep{WyaMooAcc95}.

The second and third right singular vectors indicate potential recombination hot spot locations in each of the sequence populations.    Table \ref{locations} provides a summary of all estimated recombination hot spot locations.  We see that some inferred recombination locations are consistent across each of the populations, namely the positions around 330-390 of the {\it env} sequences.   

It is also interesting to see the estimated recombination location near position 1200 of the {\it env} sequences in only the post-drug population.   There is some indication that this inferred recombination hot spot may be biologically relevant.   There is evidence of mutations that confer drug resistance that are located around position 1600 of the {\it env} sequence \citep{RayHarBla07}.   Interestingly, these drug resistant mutations were observed to have been shared across different subtypes of the HIV, where subtype identity is specified in the V3 loop region which roughly corresponds to position 950 of the {\it env} sequence.   Recombination was expected to have occurred somewhere between positions 950 and 1600 as the mechanism for sharing these resistance-conferring mutations between HIV subtypes \citep{Jensen},  and our estimated recombination location near position 1200 is located in this interval.  

RECCO provides a similar analysis in that recombination locations our methods infer are in general located in regions of low p-values, while, as expected, suggesting other regions of interest.  In this case RECCO seems to provide little more insight into the recombination structure of the {\it env} sequences, aside from further confirmatory analysis.  Further, the visual aspect of our analysis, especially the plots of the second and third right singular vectors, strongly suggest that some of the RECCO regions can be ignored altogether.  This is one really nice qualitative aspect our method can bring as supplementary to the quantitative analysis it and other methods provide- a nice, interpretable visual representation of the mutation and recombination profiles of a data set.


\begin{table}[ht]
\caption{Recombination hot spot locations inferred by our SVD method in the HIV data}
\begin{center}
\begin{tabular}{|c|c|c|c|c|}
  \cline{1-5}
  & \multicolumn{2}{|c|}{OLS} & \multicolumn{2}{|c|}{Diff} \\ \cline{1-5}
  Pre-Drug& 330 & 775 & 353 & 771 \\ \cline{1-5}
  Post-Drug & 390 & 1208 &  291 & 1208\\ \cline{1-5}
  Combined & 334 & 1055 & 358  & 1070 \\ \cline{1-5}
  \end{tabular}
\end{center}
\label{locations}
\end{table}

\section{Summary}

Our proposed approach uses the right singular vectors of a singular value decomposition to infer both the presence and location of potential recombination hot spots in genomic sequence data.  Our method is intuitive and extremely easy to implement, as the crux of our method relies only on the well-known Hamming distance and the singular value decomposition.
Our synthetic data study suggests that our method can  infer the correct number of recombination locations as well as accurately estimate the location of those recombination hotspots under a range of reasonable mutation rates.   When applied to  sequence data from  pre- and post-drug therapy HIV populations, our SVD-based method was able to estimate several putative locations of recombination hotspots.

\bibliography{writeup}

\begin{thebibliography}{9}
\providecommand{\natexlab}[1]{#1}
\providecommand{\url}[1]{\texttt{#1}}
\expandafter\ifx\csname urlstyle\endcsname\relax
  \providecommand{\doi}[1]{doi: #1}\else
  \providecommand{\doi}{doi: \begingroup \urlstyle{rm}\Url}\fi

\bibitem[Boni et~al.(2007)Boni, Posada, and Feldman]{Boni07}
Boni, Maciej~F, Posada, David, and Feldman, Marcus~W.
\newblock An exact nonparametric method for inferring mosaic structure in
  sequence triplets.
\newblock \emph{Genetics}, 176\penalty0 (2):\penalty0 1035--47, 2007.
\newblock ISSN 0016-6731.

\bibitem[Husmeier \& McGuire(2003)Husmeier and McGuire]{Husmeier03}
Husmeier, D. and McGuire, G.
\newblock {Detecting recombination in 4-taxa DNA sequence alignments with
  Bayesian hidden Markov models and Markov chain Monte Carlo.}
\newblock \emph{Mol Biol Evol}, 20\penalty0 (3):\penalty0 315--337, March 2003.
\newblock ISSN 0737-4038.

\bibitem[Jensen et~al.(2013)Jensen, Park, Braunstein, and McAuliffe]{Jensen}
Jensen, Shane~T., Park, J., Braunstein, A.F., and McAuliffe, Jon.
\newblock Bayesian hierarchical modeling of the hiv evolutionary response to
  therapy.
\newblock \emph{Journal of the American Statistical Association}, 108:\penalty0
  1230--1242, 2013.

\bibitem[Li \& Stephens(2003)Li and Stephens]{LiSte03}
Li, Na and Stephens, Matthew.
\newblock Modeling linkage disequilibrium and identifying recombination
  hotspots using single-nucleotide polymorphism data.
\newblock \emph{Genetics}, 165:\penalty0 2213--2233, 2003.

\bibitem[Maydt \& Lengauer(2006)Maydt and Lengauer]{RECCO}
Maydt, Jochen and Lengauer, Thomas.
\newblock Recco: recombination analysis using cost optimization.
\newblock \emph{Bioinformatics}, 22\penalty0 (9):\penalty0 1064--1071, 2006.
\newblock \doi{10.1093/bioinformatics/btl057}.
\newblock URL \url{http://dx.doi.org/10.1093/bioinformatics/btl057}.

\bibitem[Ray et~al.(2007)Ray, Harrison, Blackburn, Martin, Deeks, and
  Doms]{RayHarBla07}
Ray, Neelanjana, Harrison, Jessamina~E., Blackburn, Leslie~A., Martin,
  Jeffrey~N., Deeks, Steven~G., and Doms, Robert~W.
\newblock Clinical resistance to enfuvirtide does not affect susceptibility of
  human immunodeficiency virus type 1 to other classes of entry inhibitors.
\newblock \emph{Journal of Virology}, 81:\penalty0 3240--3250, 2007.

\bibitem[Schultz et~al.(2009)Schultz, Zhang, Bulla, Leitner, Korber,
  Morgenstern, and Stanke]{Schultz09}
Schultz, Anne-Kathrin, Zhang, Ming, Bulla, Ingo, Leitner, Thomas, Korber,
  Bette~T., Morgenstern, Burkhard, and Stanke, Mario.
\newblock jphmm: Improving the reliability of recombination prediction in
  hiv-1.
\newblock \emph{Nucleic Acids Research}, 37\penalty0
  (Web-Server-Issue):\penalty0 647--651, 2009.

\bibitem[Weiller(1998)]{weiller98}
Weiller, G.~F.
\newblock {Phylogenetic profiles: a graphical method for detecting genetic
  recombinations in homologous sequences}.
\newblock \emph{Mol Biol Evol}, 15\penalty0 (3):\penalty0 326--335, March 1998.

\bibitem[Wyatt et~al.(1995)Wyatt, Moore, Accola, Desjardin, Robinson, and
  Sodroski]{WyaMooAcc95}
Wyatt, R., Moore, J., Accola, M., Desjardin, E., Robinson, J., and Sodroski, J.
\newblock Involvement of the v1/v2 variable loop structure in the exposure of
  human immunodeficiency virus type 1 gp120 epitopes induced by receptor
  binding.
\newblock \emph{Journal of Virology}, 69:\penalty0 5723--5733, 1995.

\end{thebibliography}
\bibliographystyle{icml2015stylefiles/icml2015}

\end{document}